\documentclass[pre,aps,showpacs,twocolumn,floatfix]{revtex4}
\usepackage{epsfig}

\begin{document}
\title{Stability measures in metastable states with Gaussian colored noise}

\author{Alessandro Fiasconaro}
\email{afiasconaro@gip.dft.unipa.it}

\affiliation{Dipartimento di Fisica e Tecnologie Relative, Group of
Interdisciplinary Physics\footnote {http://gip.dft.unipa.it},
Universit\`a di Palermo and CNISM-INFM, Viale
 delle Scienze, I-90128 Palermo, Italy}

\author{Bernardo Spagnolo}
\affiliation{Dipartimento di Fisica e Tecnologie Relative, Group of
Interdisciplinary Physics\footnote {http://gip.dft.unipa.it},
Universit\`a di Palermo and CNISM-INFM, Viale
 delle Scienze, I-90128 Palermo, Italy}

\date{\today}
\begin{abstract}
We present a study of the escape time from a metastable state of an
overdamped Brownian particle, in the presence of colored noise
generated by Ornstein-Uhlenbeck process. We analyze the role of the
correlation time on the enhancement of the mean first passage time
through a potential barrier and on the behavior of the mean growth
rate coefficient as a function of the noise intensity. We observe
the noise enhanced stability effect for all the initial unstable
states used, and for all values of the correlation time $\tau_c$
investigated. We can distinguish two dynamical regimes characterized
by weak and strong correlated noise respectively, depending on the
value of $\tau_c$ with respect to the relaxation time of the system.
\end{abstract}

%
\maketitle

\section{Introduction}
The problem of the lifetime of a metastable state has been addressed
in a variety of areas, including first-order phase transitions,
Josephson junctions, field theory and chemical
kinetics~\cite{Tre03,Pan04}. Recent experimental and theoretical
results show that long-live metastable states are observed in
different areas of physics~\cite{Nos09,Bia09}. Experimental and
theoretical investigations have shown that the average escape time
from metastable states in fluctuating potentials presents a
nonmonotonic behavior as a function of the noise intensity with the
presence of a maximum~\cite{Man96,Agu01,Dub04}. This is the noise
enhanced stability (NES) phenomenon: the stability of metastable
states can be enhanced and the average life time of the metastable
state increases nonmonotonically with the noise intensity. This
resonance-like behavior contradicts the monotonic behavior of the
Kramers theory~\cite{Han90}. The occurrence of the enhancement of
stability of metastable states by the noise has been observed in
different physical and biological
systems~\cite{Pan04,Man96,Agu01,Dub04,Hir82,Apo97,Wac99,Pan-ova05,Sun07,Fia06,Hur06}.
Very recently NES effect was observed in an ecological
system~\cite{Rid07}, an oscillator chemical system (the
Belousov-Zhabotinsky reaction)~\cite{Yos08} and in magnetic
systems~\cite{Tra09}. Interestingly in Ref.~\cite{Yos08} the
stabilization of a metastable state due to noise is experimentally
detected and a decreasing behavior of the maximum Lyapunov exponent
as a function of the noise intensity is observed. \vspace{-0.09cm}

A generalization of the Lyapunov exponent for stochastic systems has
been recently defined in Ref.~\cite{Fia05} to complement the
analysis of the transient dynamics of metastable states. This new
measure of stability is the 'mean growth rate coefficient' (MGRC)
$\Lambda$ and it is evaluated by a similar procedure used for the
calculation of the Lyapunov exponent in stochastic
systems~\cite{Lutz-Pal96}. By linearizing the Langevin equation of
motion (see next Eq.~\ref{eq}), we consider the evolution of the
separation $\delta x(t)$ between two neighboring trajectories of the
Brownian particle starting at $x_0$ and reaching $x_F$

\begin{equation}
\label{delta} \delta \dot{x}(t)=-\frac{d^2U(x)}{dx^2} \delta x(t)=
\lambda_i(x,t) \delta x(t) \ ,
\end{equation}
and define $\lambda_i(x,t)$ as an instantaneous growth rate. We note
that, in Eq.~(\ref{delta}), $d^2U(x)/dx^2$ is calculated onto the
noisy trajectory $x[\xi(t)]$~\cite{Fia05}. The growth rate
coefficient $\Lambda_i$ (for the $i_{th}$ noise realization), is
then defined as the long-time average of the instantaneous
$\lambda_i$ coefficient over
$\tau(x_0,x_F)$~\cite{Fia05,Lutz-Pal96,Bof02}
\begin{equation}
\label{lyap} \Lambda_i = \frac{1}{\tau(x_0,x_F)}
\int_{0}^{\tau(x_0,x_F)} \lambda_i(x,s) ds.
\end{equation}
In the limit $\tau(x_0,x_F) \to \infty$, Eq.~(\ref{lyap}) coincides
formally with the definition of the maximum Lyapunov exponent, and
therefore, the $\Lambda_i$ coefficient has the meaning of a
finite-time Lyapunov exponent. This quantity is useful to
characterize a transient dynamics in nonequilibrium dynamical
systems~\cite{Yos08,Fia05}. The mean growth rate coefficient
$\Lambda$ is then defined as the ensemble average of the growth rate
coefficient $\Lambda_i$
\begin{equation}
\Lambda =\;\;< \Lambda_i >
\end{equation}
over the noise realizations. The mean growth rate coefficient has a
nonmonotonic behavior as a function of the noise intensity for
Brownian particles starting from unstable initial
positions~\cite{Fia05}. This nonmonotonicity with a minimum
indicates that $\Lambda$ can be used as a new suitable measure or
signature of the NES effect.

The inclusion of realistic noise sources, with a finite correlation
time, impacts both the stationary and the dynamic features of
nonlinear systems. For metastable thermal equilibrium systems it has
been demonstrated that colored thermal noise can substantially
modify the crossing barrier process~\cite{Han90}. A rich and
enormous literature on escape processes driven by colored noise was
produced in the $80$'s~\cite{Rub88,Ram91,Sanc89}. More recently many
papers investigated the role of the correlated noise on different
physical systems~\cite{Yos07,Chi08,Kam08,Val05,Fia-b05,Sen09}, which
indicates a renewed interest in the realistic noise source effects.

In this work we present a study of the average decay time of an
overdamped Brownian particle subject to a cubic potential with a
metastable state. We focus on the role of different unstable initial
conditions and of colored noise in the average escape time. The
effect of the correlation time $\tau_c$ on the transient dynamics of
the escape process is related to the characteristic time scale of
the system, that is the relaxation time inside the metastable state
$\tau_r$. For $\tau_c < \tau_r$, the dynamical regime of the
Brownian particle is close to the white noise dynamics. For $\tau_c
> \tau_r$, we obtain: (i) a big shift of the increase of the average
escape times towards higher noise intensities; (ii) an enhancement
of the value of the average escape time maximum with a broadening of
the NES region in the plane ($\tau, D$), which becomes very large
for high values of $\tau_c$; (iii) the shift of the peculiar initial
position $x_c$ (towards lower values), found in our previous
studies~\cite{Dub04, Fia05}, which separates the set of the initial
unstable states producing divergency, for $D$ tending to zero, from
those which give only a nonmonotonic behavior of the average escape
time; (iv) the entire qualitative behaviors (i-iii) can be applied
to the standard deviation of the escape time; (v) the shift of the
minimum values in the curves of the mean growth rate coefficient
$\Lambda$; (vi) trend to the disappearance of the minimum in the
curves of $\Lambda$, with a decreasing monotonic behavior for
increasing $\tau_c$; (vii) trend to the disappearance of the
divergent dynamical regime in $\tau$, with increasing $\tau_c$. The
paper is organized as follows. In the next section we introduce the
model. In the third section we show the results and in the final
section we draw the conclusions.

\section{The model}

\begin{figure}[htbp]
\begin{center}
\includegraphics[height=7cm,angle=-90]{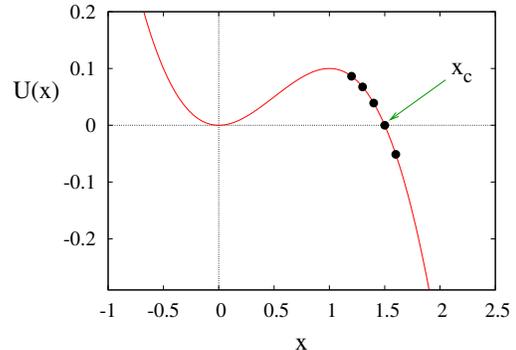}
\caption{(Color online) The cubic potential $U(x) = a x^2 - b x^3$
with the various initial positions investigated (dots), namely $x_o
= 1.2, 1.3, 1.4, 1.5, 1.6$. The parameters of the potential are: $a
= 0.3, b = 0.2$. For the white noise case, $x_c = 1.5$ is the
critical initial position which separates the set of the initial
unstable states producing divergency, for $D$ tending to zero, from
those which give only a nonmonotonic behavior of the average escape
time~\cite{Agu01,Fia05}.} \label{1-fpot}
\end{center}%
\end{figure}

The starting point of our study is the Langevin equation
\begin{equation}
\dot{x}=-\frac{\partial U(x)}{\partial x} + \eta(t)
\label{eq}
\end{equation}
where $\eta(t)$ is the Ornstein-Uhlenbeck process
\begin{equation}
d\eta=-k \eta dt + k \sqrt{D} \; d W(t)
\label{eqou}
\end{equation}
and $W(t)$ is the Wiener process with the usual statistical
properties: $\langle \xi(t) \rangle = 0$ and $\langle
\xi(t)\xi(t+\tau) \rangle = \delta (\tau)$. The system of
Eqs.~(\ref{eq}) and~(\ref{eqou}) represents a two-dimensional
Markovian process, which is equivalent to a non-Markovian Langevin
equation driven with additive Gaussian correlated noise, with
$\eta(t)$ obeying the following statistical properties $\langle
\eta(t) \rangle = 0$ and $\langle \eta(t) \eta(t+\tau) \rangle =
(kD/2) e^{-k\tau}$, for $t \to \infty$ and $\eta(0) = 0$. Here
$1/k=\tau_c$ is the correlation time of the process. The integration
of Eq.~(\ref{eqou}) yields in the limit $\tau_c \to 0$ the white
noise term
 \begin{equation}
\lim_{\tau_c \to 0} \eta(t) = 2\sqrt{D} \int_0^t \lim_{\tau_c\to 0}
\frac{e^{-(t-t')/\tau_c}}{2\tau_c} dW(t') = \sqrt{D} \xi(t),
\label{etlim}
\end{equation}
and the stationary correlation function of the Ornstein-Uhlenbeck
process gives in the limit $\tau_c \to 0$ the correlation function
of the white noise: $\lim_{\tau_c \to 0} \langle \eta(t)
\eta(t+\tau) \rangle = D\delta(\tau)$. The potential $U(x)$ used in
Eq.~(\ref{eq}) is $U(x)=ax^2 - bx^3$, with $a=0.3$, $b=0.2$. The
potential profile has a local stable state at $x=0$  and an unstable
state at $x=1$ (see Fig.~\ref{1-fpot}). The relaxation time for the
metastable state at $x=0$ is $\tau_r = \left[\frac{d^2
U(x)}{dx^2}\right]_{x=0} = 2 a$, which is the characteristic time
scale of our system. For our potential profile we have $\tau_r =
0.6$.
\section{Results}
\begin{figure}[htbp]
\centering
\includegraphics[width=8.5cm]{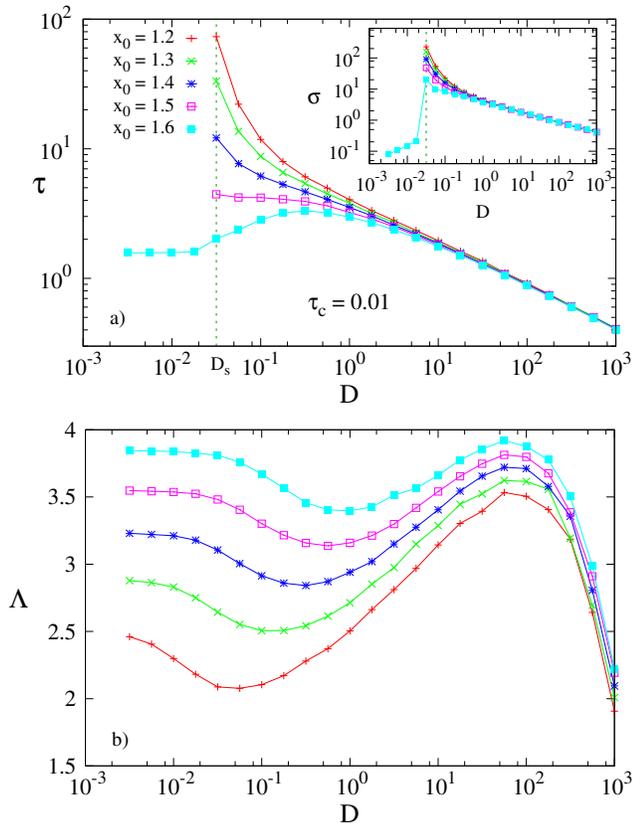}
\caption{(Color online) Panel a: Log-Log plot of the mean first
passage time $\tau$ as a function of noise intensity $D$ in the case
of correlated noise with $\tau_c = 0.01$, for the four initial
positions investigated (see Fig.~\ref{1-fpot}). Inset: the related
standard deviation as a function of the noise intensity $D$. The
dotted straight line at $D = D_s$ separates the simulation data
representing the Brownian particles escaped within the maximum
simulation time $T_{max}$ for $D > D_s$, from those representing the
particles partially trapped within the well for a time greater or
equal to $T_{max}$ for $D < D_s$. Panel b: Mean growth rate
coefficient $\Lambda$ as a function of the noise intensity $D$, with
the same initial positions of Fig.~\ref{1-fpot}.} \label{2-nes}
\end{figure}
\begin{figure}[htbp]
\centering
\includegraphics[width=8.5cm]{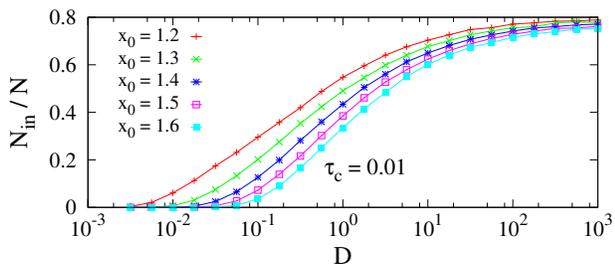}
\caption{(Color online) Semi-Log plot of the fraction of particles
$N_i/N$ reaching the threshold position $x_t = 0.5$ into the
potential well, within the $T_{max}$, as a function of noise
intensity $D$. This threshold position $x_t$ corresponds to the flex
point of the potential, where the instantaneous growth rate
$\lambda_i(x,t)$ is equal to zero. The correlation time of the noise
is $\tau_c = 0.01.$, with the same initial conditions of
Fig.~\ref{1-fpot}.} \label{3-n_i}
\end{figure}
The calculations of the average escape time as a function of the
colored noise intensity have been performed by averaging over
\textbf{$N = 20,000$ }realizations the numerical solution of the
stochastic differential equation~(\ref{eq}). The absorbing boundary
for the escape process is put on $x_F = 20$, and the maximum
simulation time is $T_{max} = 10,000~a.u.$. For all the initial
unstable states (see Fig.~\ref{1-fpot}) and all the correlation
times considered we find an enhancement of the mean first passage
time (MFPT) $\tau$ with respect to the deterministic time.
\begin{figure*}[htbp]
\centering
\includegraphics[height=17.4cm,angle=-90]{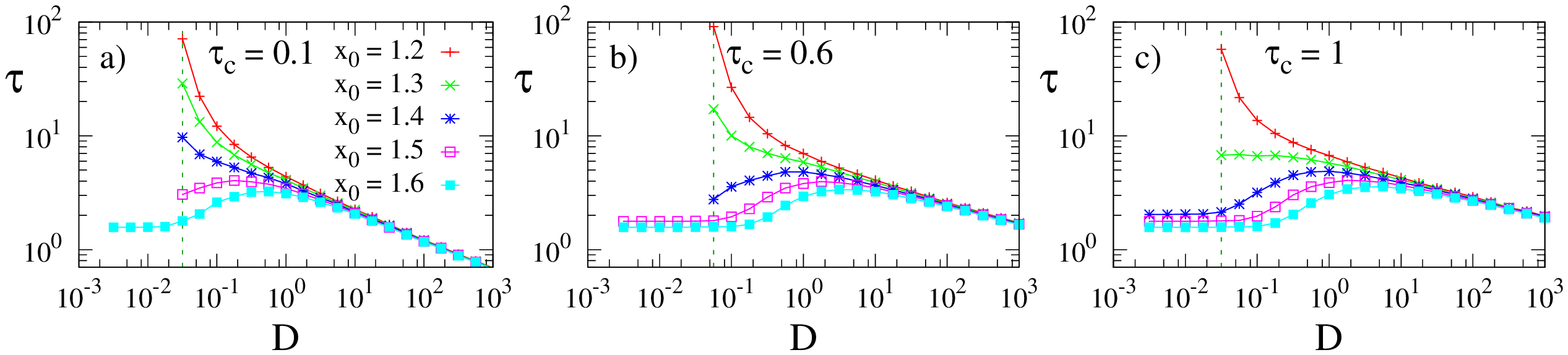}
\caption{(Color online) Log-Log plot of the MFPT $\tau$ as a
function of noise intensity $D$ for the same initial positions of
Fig.~\ref{1-fpot} and for different values of the correlation times
$\tau_c$, namely $\tau_c = 0.1, 0.6, 1$, corresponding respectively
to the weak, intermediate and strong colored noise dynamical regime.
The dotted straight line at $D = D_s$ separates the noise values for
which all the Brownian particles escape ($D > D_s$), from those for
which the particles are partially trapped into the well within the
$T_{max}$ ($D < D_s$).} \label{4-tau_c}
\end{figure*}
\begin{figure}[htbp]
\centering
\includegraphics[height=8.0cm,angle=-90]{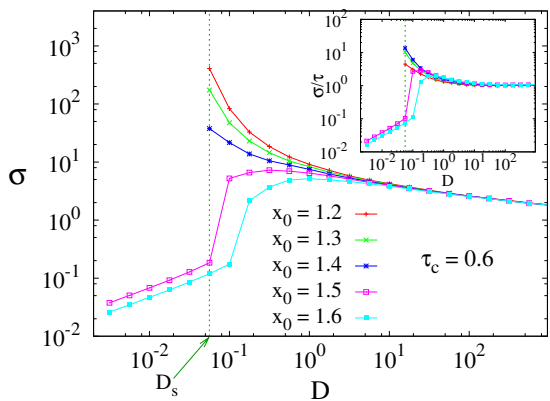}
\caption{(Color online) Log-Log plot of the standard deviation
$\sigma$ as a function of the noise intensity $D$ for $\tau_c = 0.6$
and the same initial positions of Fig.~\ref{1-fpot}. The dotted
straight line indicates the value of the noise intensity $D_s$ which
separates the simulation data representing the trapped Brownian
particles from those escaped within $T_{max}$. Inset: Log-Log plot
of the ratio $\sigma/\tau$ as a function of noise intensity $D$.}
\label{5-sigma}
\end{figure}
\begin{figure*}[htbp]
\centering
\includegraphics[height=17.4cm,angle=-90]{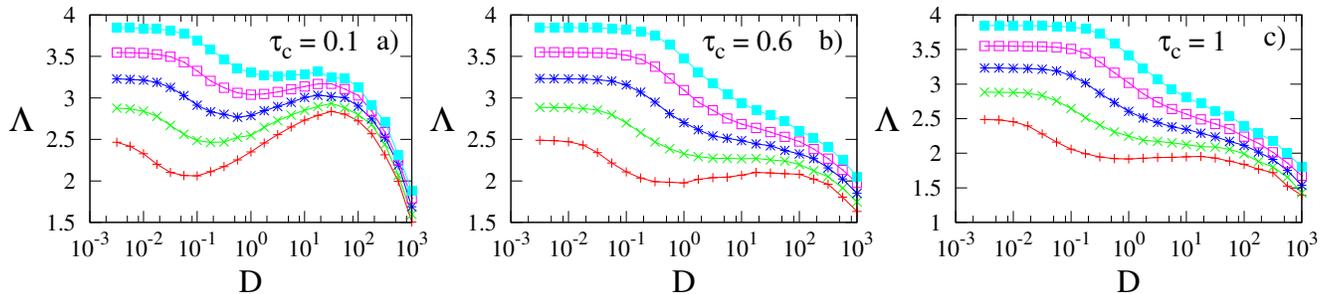}
\caption{(Color online) Semi-Log plot of mean growth rate
coefficient $\Lambda$ as a function of noise intensity $D$ for the
same initial positions $x_o$ and the same values of the correlation
times $\tau_c$ of Fig.~\ref{4-tau_c}, namely $\tau_c = 0.1, 0.6, 1$,
corresponding respectively to the weak, intermediate and strong
colored noise dynamical regime.} \label{6-lya_c}
\end{figure*}
\begin{figure}[htbp]
\centering
\includegraphics[width=8.5cm]{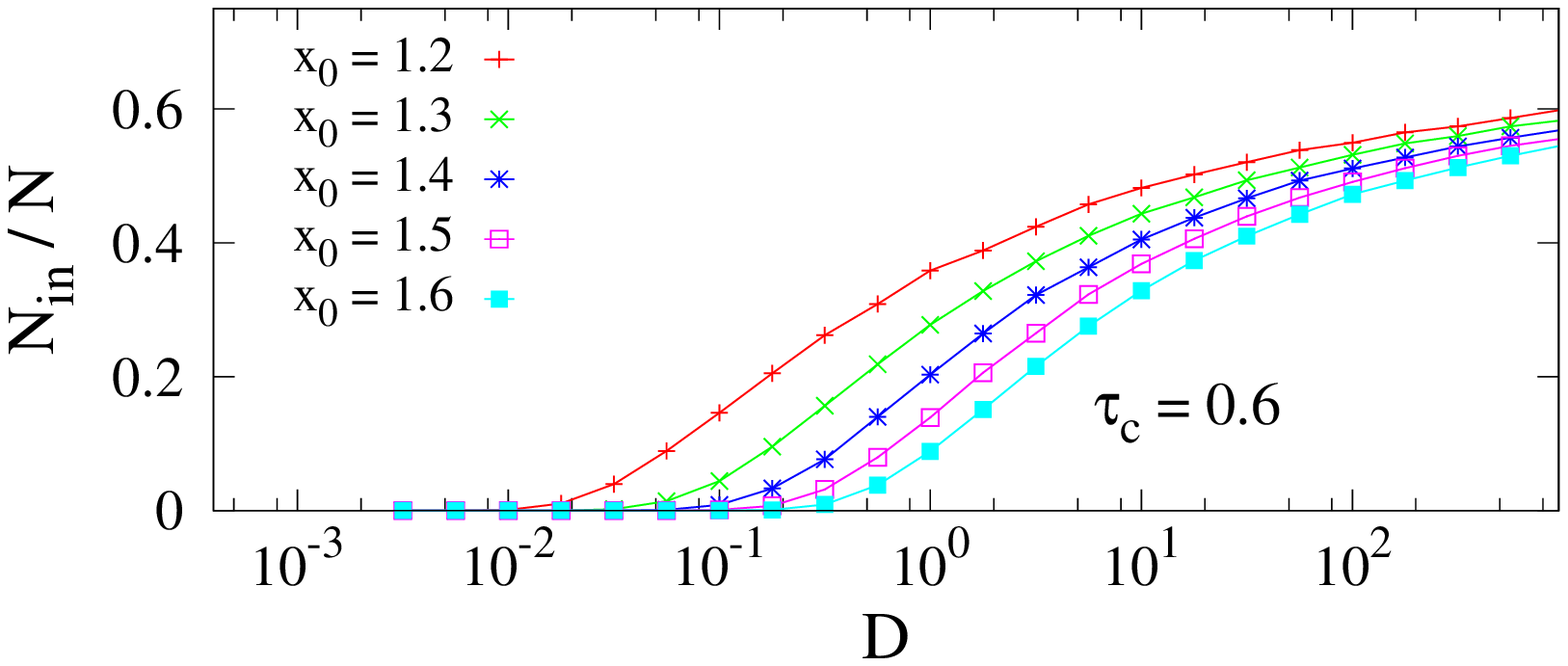}
\caption{(Color online) Semi-Log plot of the fraction of particles
$N_i/N$ reaching the threshold position $x_t = 0.5$ into the
potential well, within the $T_{max}$, as a function of noise
intensity $D$. This threshold position $x_t$ corresponds to the flex
point of the potential, where the instantaneous growth rate
$\lambda_i(x,t)$ is equal to zero. The correlation time of the noise
is $\tau_c = 0.6$, with the same initial conditions of
Fig.~\ref{1-fpot}.} \label{7-n_i_c}
\end{figure}

In Fig.~\ref{2-nes} the calculations performed with low colored
noise ($\tau_c=0.01$) for the mean first passage time $\tau$ and the
mean growth rate coefficient $\Lambda$ are shown. We see as
signatures of the NES effect a maximum in the curve of $\tau$ and a
minimum in that of $\Lambda$. In the inset of Fig.~\ref{2-nes}a the
standard deviation of the first passage time as a function of noise
intensity is reported. We note that the behaviors of $\tau$ and
$\Lambda$ in this low colored noise regime ($\tau_c = 0.01$)
coincides with those obtained in the white noise case~\cite{Fia05}.
Moreover by comparing the theoretical predictions of $\tau$ (see
Eq.~(3) of Ref.~\cite{Fia05}) with direct numerical simulations of
the Langevin equation, a very good agreement is obtained (see Fig.~3
of Ref.~\cite{Fia05}). In Fig.~\ref{3-n_i} the semi-Log plots of the
fraction of particles $N_i/N$ reaching the threshold position $x_t =
0.5$ into the potential well, within the $T_{max}$, as a function of
noise intensity $D$, with the same initial conditions of
Fig.~\ref{1-fpot}, are shown. This threshold position $x_t$
corresponds to the concavity change of the potential and is
considered for this reason as a reference indicator for the
effective entrance of the particle into the well. It is possible to
observe that for very low noise intensity none particle enters into
the well within the $T_{max}$ considered, and the estimation of the
stability measures take their deterministic values. We note that the
behavior of the mean growth rate coefficient as a function of the
noise intensity is strongly affected by the characteristic potential
shape of a metastable state. The curves shown in Fig.~\ref{3-n_i}
clarify the behavior of $\Lambda$ in the limit of $D \rightarrow 0$.
In fact the position $x_t = 0.5$ is the flex point of the potential,
where the instantaneous growth rate $\lambda_i(x,t)$ is equal to
zero. We see that for low noise intensities the fraction $N_i/N$
goes to zero, producing an increasing behavior of the MGRC (see
Fig.~\ref{2-nes}b).

The behaviors of the MFPTs as a function of the noise intensity $D$
with other values of $\tau_c$ are shown in Fig.~\ref{4-tau_c}. We
clearly observe two dynamical regimes depending on the value of
$\tau_c$ with respect to the relaxation time of the system ($\tau_r
= 0.6$): (a) weak colored noise ($0< \tau_c <\tau_r$) and (b) strong
colored noise ($\tau_c > \tau_r$). By observing Fig.~\ref{4-tau_c}a
($\tau_c = 0.1$) we can see that the qualitative behavior of MFPT
shown in Fig.~\ref{2-nes}a is recovered. In the weak color noise
regime we can still observe the divergent behavior of MFPTs for
$x_{max}<x_0<x_c$ and a non monotonic behavior for $x_0 \ge x_c$,
with $x_c = 1.5$. By increasing the value of the correlation time
($\tau_c \ge \tau_r$) we observe a large displacement of the maximum
of MFPT towards higher values of noise intensity and a shift of the
peculiar initial position $x_c$ towards lower values. For $\tau_c =
\tau_r = 0.6$, $x_c^* \simeq 1.4$, and for $\tau_c = 1$, $x_c^*
\simeq 1.3$ (see Figs.~\ref{4-tau_c}b and~\ref{4-tau_c}c), where
$x_c^*$ is the peculiar initial position of the Brownian particle in
the presence of colored noise. We note that $x_c = 1.5$ is a fixed
value for white noise case~\cite{Fia05}, while the position $x_c^*$
is a variable quantity for colored noise and it is depending on the
value of the correlation time of the noise. Moreover, we observe a
broadening of the NES region, which becomes very large for high
values of the correlation time $\tau_c$. The NES region is the area
where enhanced stability of a metastable state is observed. In other
words it is the area under each curve of $\tau$ \emph{vs} $D$ (see
Figs.~\ref{2-nes}a and~\ref{4-tau_c}), where the values of $\tau$
are greater than the deterministic dynamical time related to the
particular initial position investigated (see also Fig.~1 in
\emph{Mantegna and Spagnolo}, 1998, Ref.~\cite{Man96}).
\vspace{-0.09cm}

The asymmetry of the potential profile with respect to the $x$
coordinate makes more effective the correlation of the noise for
Brownian particles moving from left to right. This means that, at
very low noise intensities of the colored noise, the particles
inside the potential well will escape more easily with respect to
the white noise case. Therefore, the trapping effect, which is
responsible for the divergent behavior for any initial unstable
state within the range $x_{max} < x_o < x_c$ will happen in a
restricted range of initial positions, that is $x_{max} < x_o <
x_c^*$ with $x_c^* < x_c$. Specifically this peculiar position
$x_c^*$ is shifted towards decreasing values of the $x$ coordinate
for increasing correlation time $\tau_c$ of the noise source. In
Fig.~\ref{2-nes}a and all panels of Fig.~\ref{4-tau_c} the dotted
straight line at $D = D_s$ separates the simulation data
representing the Brownian particles escaped within the maximum
simulation time $T_{max}$ for $D > D_s$, from those representing the
particles partially trapped within the well for a time greater or
equal to $T_{max}$ for $D < D_s$. This means that the simulation
data obtained for $D < D_s$ underestimate the real data in the
divergent dynamical regime. In fact if we prolong the maximum
simulation time $T_{max}$ we obtain more approximate values for
$\tau$ and $\sigma$ and the divergent behavior will be visible at
lower noise intensities. As a consequence $D_s$ will be shifted
towards lower values.

For high values of the noise intensity all the plots show a
monotonic decrease behavior as a function of noise intensity
collapsing in a unique curve. Moreover the slope of this limit curve
becomes flatter by increasing the correlation time. This means that
the NES effect involves more and more orders of magnitude of the
noise intensity. The effect of the colored noise is therefore to
delay the escape process or in other words to enhance more and more
the stability of the metastable state for increasing values of the
noise intensity.

In Fig.~\ref{5-sigma} the standard deviation $\sigma$ of the first
passage time distribution for $\tau_c = 0.6$ is shown. We see a huge
increase of the $\sigma$ for low values of noise intensity,
demonstrating a strong enlargement of the distribution when the
particle feels a noise intensity comparable with the height of
potential barrier. Similarly to the MFPTs, \textit{color} induces a
shift in the divergent behavior of $\sigma$. The relative measure of
the width with respect to the mean value is shown in the inset of
Fig.~\ref{5-sigma}, where the ratio $\sigma / \tau$ is plotted. This
ratio reveals a nonmonotonic behavior with a minimum, demonstrating
the existence of a noise intensity for which the width of the first
passage time distribution is the minimum related to its mean. In
other words this value corresponds to a maximum of precision in the
measure of $\tau$. This optimal noise intensity is shifted toward
high noise values by increasing $\tau_c$.

The behavior of the mean growth rate coefficient $\Lambda$ as a
function of the noise intensity $D$ for different values of the
noise correlation time is shown in Fig.~\ref{6-lya_c}. In the weak
color noise regime we observe a nonmonotonic behavior with a minimum
for all the initial positions investigated with a shift in the
position of the minimum towards higher noise intensities. In the
strong color regime the minimum, which represents a trapping
phenomenon for a finite time, is visible for the divergent behavior
of MFPTs for $x_{max}<x_0<x_c^*$ and it is shifted towards higher
noise intensities by increasing the correlation time. For initial
positions $x_0 \ge x_c^*$, the minimum tends to disappear, but at
the same time the $\Lambda$ parameter decreases monotonically with
increasing noise intensity, showing a trapping phenomenon at higher
noise intensities. This trend to the disappearance of the minimum in
the curves of $\Lambda$, corresponds to the restricted range of the
initial positions for which we observe a divergent behavior of
$\tau$, that is to the trend of disappearance of this divergent
behavior. We note that the behaviour of $\Lambda$ as a function of
the noise intensity $D$ obtained in our analysis is in qualitative
agreement with that obtained by the experimental investigation of
the stabilization of a metastable state in an oscillatory chemical
system (the Belousov-Zhabotinsky reaction)~\cite{Yos08}.
Specifically the decreasing behavior of the maximum Lyapunov
exponent of Fig.~2 of Ref.~\cite{Yos08} is in qualitative good
agreement with the behavior of the curves in Fig.~\ref{6-lya_c}b and
Fig.~\ref{6-lya_c}c. This could be ascribed to the correlation time
always present in noise sources used in any experimental set-up.

In Fig.~\ref{7-n_i_c} we report, for all the initial positions
investigated and for $\tau_c = 0.6$, the semi-Log plot of the
fraction of particles $N_i/N$ entering into the potential well up to
the position $x_t=0.5$, within the $T_{max}$, as a function of noise
intensity $D$. At very low noise intensities and for increasing
values of the correlation time $\tau_c$, the particles have
difficulty to enter into the potential well, within the $T_{max}$
considered, shifting the entrance statistics towards higher values
of the noise intensity.

\section{Conclusions}
In this work we analyzed the effect of the colored noise, generated
by an Ornstein-Uhlenbeck process, on the enhancement of the mean
first passage time in a cubic potential with a metastable state and
on the minimum of the mean growth rate coefficient as a function of
the noise intensity. We analyze different initial unstable states.
We obtain NES effect for all the initial positions investigated and
an enhancement of the NES region for increasing values of
correlation times. The results obtained for a particle moving in a
cubic potential are quite general, because we always obtain NES
effect when a particle is initially located just to the right of a
local potential maximum and next to a metastable state, in the
escape region.

In experiments real noise sources are correlated with a finite
correlation time. As a consequence the NES effect can be observed at
higher noise intensities with respect to the idealized white noise
case. The enhancement and the shift of the NES region, towards
higher values of the noise intensity, allows to reveal
experimentally the NES effect only by using a suitable correlation
time $\tau_c$ in the noise source.

This work was supported by MIUR.

\end{document}